\documentclass[twocolumn,twoside,slac_two]{revtex4}
\usepackage{graphicx}
\usepackage{fancyhdr}
\begin{document}

\title{New Physics from Flavour}


\author{M.~Bona}
\affiliation{CERN, CH-1211 Geneva 23, Switzerland}
\author{M.~Ciuchini}
\affiliation{Dipartimento di Fisica, Universit\'a di Roma Tre and INFN Roma III, Italy}
\author{E.~Franco}
\affiliation{Dipartimento di Fisica, Universit\'a di Roma ``La Sapienza'' and INFN Roma, Italy}
\author{V.~Lubicz}
\affiliation{Dipartimento di Fisica, Universit\'a di Roma Tre and INFN Roma III, Italy}
\author{G.~Martinelli}
\affiliation{Dipartimento di Fisica, Universit\'a di Roma ``La Sapienza'' and INFN Roma, Italy}
\author{F.~Parodi}
\affiliation{Dipartimento di Fisica, Universit\'a di Genova and INFN Genova, Italy}
\author{M.~Pierini}
\affiliation{CERN, CH-1211 Geneva 23, Switzerland}
\author{C.~Schiavi}
\affiliation{Dipartimento di Fisica, Universit\'a di Genova and INFN Genova, Italy}
\author{L.~Silvestrini}
\affiliation{Dipartimento di Fisica, Universit\'a di Roma ``La Sapienza'' and INFN Roma, Italy}
\author{V.~Sordini}
\affiliation{Laboratoire de l'Acc\'el\'erateur Lin\'eaire, IN2P3-CNRS and Universit\'e de Paris-Sud, Orsay Cedex, France}
\author{A.~Stocchi}
\affiliation{Laboratoire de l'Acc\'el\'erateur Lin\'eaire, IN2P3-CNRS and Universit\'e de Paris-Sud, Orsay Cedex, France}
\author{C.~Tarantino}
\affiliation{Dipartimento di Fisica, Universit\'a di Roma Tre and INFN Roma III, Italy}
\author{V.~Vagnoni (corresponding author)}
\affiliation{INFN Bologna, Italy}

\begin{abstract}
The UT{\it{fit}} Collaboration has produced several
analyses in the context of flavour physics both within and
beyond the Standard Model.
In this paper we present updated results for the Standard Model analysis of the 
Unitarity Triangle using the latest experimental and lattice QCD inputs, as well as an update
of the Unitarity Triangle analysis in a scenario beyond the Standard Model. Combining all
available experimental and theoretical information on
$\Delta F=2$ processes and using a model-independent parameterization,
we extract the allowed New Physics contributions in the
$K^0$, $D^0$, $B_d$, and $B_s$ sectors. We observe a departure
of the $B_s$ mixing phase from the Standard Model expectation
with a significance of about $3\sigma$.
\end{abstract}

\maketitle

\thispagestyle{fancy}


\section{Introduction}
The UT{\it{fit}} Collaboration~\cite{utfitsite} aims to
determine the coordinates $\bar\rho$ and $\bar\eta$ of the apex of the
Unitarity Triangle (UT), and in general the elements of the CKM
matrix~\cite{CKM} in the Standard Model (SM). Nowadays the SM analysis
includes many experimental and theoretical results, such as predictions for several flavour
observables and measurements of hadronic parameters which can be compared with
the lattice QCD predictions~\cite{utfitsm}.
More recently, the UT analysis has been extended beyond the SM, allowing for a
model-independent determination of $\bar\rho$ and $\bar\eta$ --- assuming
negligible New Physics (NP) contributions to
tree-level processes --- and a simultaneous evaluation of the size of NP contributions
to $\Delta F=2$ amplitudes compatible with the flavour data~\cite{utfitnp,utfitbetas}.
Recently, the NP analysis has been expanded to include an effective field theory
study of the allowed NP contributions to $\Delta F=2$ amplitudes. This allows one to
put model-independent bounds on the NP energy scale associated to flavour- and
CP-violating phenomena~\cite{utfiteft}.

\begin{table}
  \caption{Input parameters used in the SM UT fit. The first
error corresponds to the width of a Gaussian, while the second one, whenever present, is the half width
of a uniform distribution. The two
distributions are then convolved to obtain the final one. Entries marked with 
$(\dagger)$ are only indicative of the $68\%$ probability ranges, as the full
experimental likelihood has
actually been used to obtain the prior distributions for these parameters.
Entries without errors are considered as constants in the fit.}
  \label{tab:inputs}
  \begin{tabular}{|ll|}
    \hline
      $\alpha_s(M_Z)$ & $0.119\pm 0.003$\\
      $G_F$ & $1.16639\cdot 10^{-5}$ GeV$^{-2}$\\
      $M_W$ & $80.425$ GeV\\ 
      $M_Z$ & $91.1876$ GeV\\
      ${\bar m}_t({\bar m}_t)$ & $(162.8\pm 1.3)$ GeV  \\
      ${\bar m}_b({\bar m}_b)$ & $(4.21\pm 0.08)$ GeV \\
      ${\bar m}_c({\bar m}_c)$ & $(1.3\pm 0.1)$ GeV \\
      ${\bar m}_s(2$ GeV$)$ & $(105\pm 15)$ MeV \\
    \hline
      $M_{Bd}$ & $5.279$ GeV\\
      $M_{Bs}$ & $5.375$ GeV\\
%      $\tau_{D}$ & $(0.4101\pm 0.0015)$ ps \\
      $\tau_{B_d}$ & $(1.527\pm 0.008)$ ps \\
      $\tau_{B^+}$ & $(1.643\pm 0.010)$ ps \\
      $\tau_{B_s}$ & $(1.39\pm 0.12)$ ps \\
    \hline
      $|V_{cb}|$ (exclusive) & $(3.92\pm 0.11)\cdot 10^{-2}$\\
      $|V_{cb}|$ (inclusive) & $(4.168\pm 0.039\pm 0.058)\cdot 10^{-2}$\\
      $|V_{ub}|$ (exclusive) & $(3.5\pm 0.4)\cdot 10^{-3}$\\
      $|V_{ub}|$ (inclusive) & $(4.00\pm 0.15\pm 0.40)\cdot 10^{-3}$\\
    \hline
      $\varepsilon_K$ & $(2.232\pm 0.007)\cdot 10^{-3}$\\
      $M_K$ & $497.648$ MeV\\
      $f_K$ & $160$ MeV \\
      ${\hat B}_K $ & $0.75\pm 0.07$\\
    \hline
      $\Delta m_{d}$ & $(0.507\pm 0.005)$ ps$^{-1}$\\
      $\Delta m_{s}$ & $(17.77\pm 0.12)$ ps$^{-1}$\\
      $f_{B_s} \sqrt {{\hat B}_{B_s}}$ & $(270\pm 30)$ MeV\\
      $\xi=f_{B_s} \sqrt {{\hat B}_{B_s}}/f_{B_d} \sqrt {{\hat B}_{B_d}}$ & $1.21\pm 0.04$\\
    \hline
      $\lambda$ & $0.2258\pm 0.0014$\\
      $\alpha (^\circ)$ & $92\pm 8~(\dagger)$ \\
      $\sin2\beta$ & $0.668\pm 0.028~(\dagger)$ \\
      $\cos2\beta$ & $0.88\pm 0.12~(\dagger)$\\
      $\gamma (^\circ)$ & $(80\pm 13)\cup (-100\pm 13)~(\dagger)$ \\
      $(2\beta+\gamma) (^\circ)$ & $ (94\pm 53)\cup (-90\pm 57)~(\dagger)$\\
    \hline
      $BR(B^+\to\tau^+\nu_\tau)$ & $(1.12\pm 0.45)\cdot 10^{-4}~(\dagger)$ \\
      $f_{B_d}$ & $(200\pm 20)$ MeV\\
    \hline
  \end{tabular}
\end{table}

In these proceedings we present a preliminary update of our UT analysis in the SM,
including a set of fit predictions and a study of the compatibility between
the fit results and some of the most interesting experimental constraints.
The main difference with respect to previously published results comes from the
use of an updated set of lattice QCD results~\cite{lubicz} and of some constraints
($\bar m_t$, $\alpha$, $\gamma$, $|V_{ub}|$) updated to the latest
available measurements. 
We also show an update of the analysis beyond the SM, with particular emphasis on NP
contributions to the $B_s$ mixing phase, where we observe a significant discrepancy with
respect to the SM prediction.

\begin{figure}
\centering
\includegraphics[width=.48\textwidth]{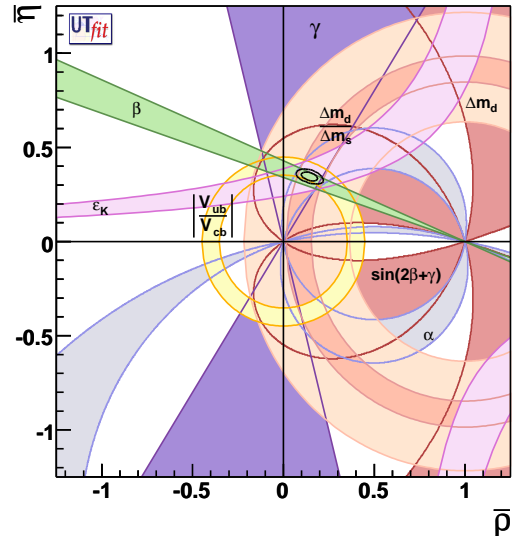}
\caption{Result of the SM fit. The contours show the $68\%$ and $95\%$ probability
regions selected by the fit in the $\bar\rho$--$\bar\eta$ plane.
The $95\%$ probability regions selected by the single constraints are also shown.}
\label{fig:all}
\end{figure}

\begin{table}
  \caption{Results of the SM fit obtained using the experimental constraints
discussed in the text. We quote the $68\%$ [$95\%$] probability
ranges.}
  \label{tab:results}
  \begin{tabular}{|lll|}
    \hline
      $\lambda$ & $0.2259\pm 0.0015$ & [$0.2228,0.2288$]\\
      $A$ & $0.809\pm 0.013$ & [$0.783,0.835$]\\
      $\bar\rho$ & $0.155\pm 0.022$ & [$0.112,0.197$]\\
      $\bar\eta$ & $0.342\pm 0.014$ & [$0.316,0.370$]\\
    \hline
      $R_b$ & $0.377\pm 0.013$ & [$0.352,0.403$]\\
      $R_t$ & $0.911\pm 0.022$ & [$0.866,0.953$]\\
      $\alpha (^\circ)$ & $92.1\pm 3.4$ & [$85.7,99.0$]\\
      $\beta  (^\circ)$ & $22.0\pm 0.8$ & [$20.5,23.7$]\\
      $\gamma (^\circ)$ & $65.6\pm 3.3$ & [$58.9,72.1$]\\
    \hline
      $|V_{cb}|\cdot 10^{2}$ & $4.125\pm 0.045$ & [$4.04,4.21$]\\
      $|V_{ub}|\cdot 10^{3}$ & $3.60\pm 0.12$ & [$3.37,3.85$]\\
      $|V_{td}|\cdot 10^{3}$ & $8.50\pm 0.21$ & [$8.07,8.92$]\\
      $|V_{td}/V_{ts}|$ & $0.209\pm 0.005$ & [$0.199,0.219$]\\
    \hline
      Re$\lambda_t\cdot 10^{3}$ & $-0.32\pm 0.01$ & [$-0.34,-0.30$]\\
      Im$\lambda_t\cdot 10^{5}$ & $13.5\pm 0.5$ & [$12.4,14.6$]\\
    \hline
      $J_{CP}\cdot 10^{5}$ & $2.98\pm 0.12$ & [$2.75,3.22$]\\
    \hline
      $\Delta m_{s} (\mathrm{ps}^{-1})$ & $17.75\pm 0.15$ & [$17.4,18.0$]\\
      $\sin2\beta_s$ & $0.0365\pm 0.0015$ & [$0.0337,0.0394$]\\
    \hline
  \end{tabular}
\end{table}

\section{The Unitarity Triangle analysis in the Standard Model}

In the UT analysis we combine the available theoretical
and experimental information relevant to determine $\bar\rho$ and $\bar\eta$.
To this end, we use a Bayesian approach as described in
ref.~\cite{dago}.
The theoretical and experimental input values and errors are collected in
Table~\ref{tab:inputs}.

The results of the SM fit are shown in Table~\ref{tab:results}, while
the $\bar\rho$--$\bar\eta$ plane can be found in Figure~\ref{fig:all}, where
the $68\%$ and $95\%$ probability regions are plotted together
with the $95\%$ regions selected by the single constraints. It is quite remarkable that
the overall picture looks very consistent. The parameters $\bar\rho$ and $\bar\eta$ are determined
in the SM with a relative errors of $14\%$ and $4\%$ respectively.

\begin{figure}
\centering
\includegraphics[width=.48\textwidth]{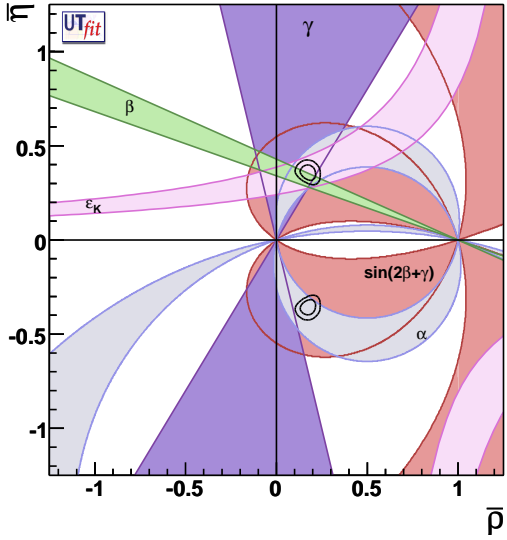}
\includegraphics[width=.48\textwidth]{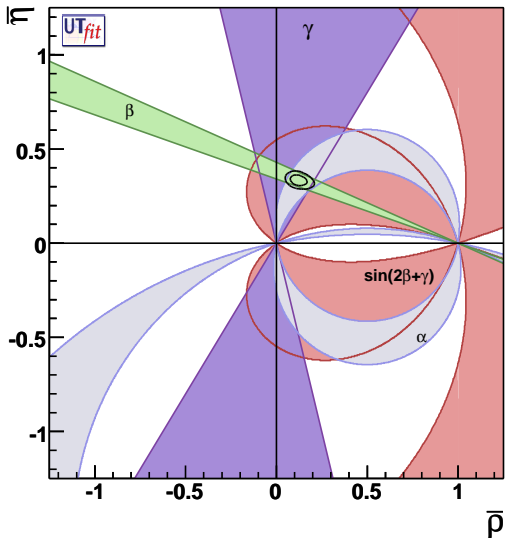}
\caption{Constraints in the $\bar\rho$--$\bar\eta$ plane from the
measurement of CP-conserving observables only (left).
Constraints in the $\bar\rho$--$\bar\eta$ plane from the
measurement of the angles of the UT only (right). }
\label{fig:angles}
\end{figure}

Within the precision of $\sim 5$--$10\%$, the CKM mechanism of the SM
is able to describe pretty well the violation of the CP symmetry. In addition,
flavour-changing CP-conserving and CP-violating processes select compatible 
regions in the $\bar\rho$--$\bar\eta$ plane, as predicted by the three-generation 
unitarity. This is illustrated on the left side of fig.~\ref{fig:angles},
while on the right side we show the constraining power of the CP-violating 
observables (namely the UT angles) in the $B_d$ sector only.

The results of the fit are displayed in Table~\ref{tab:results}.
In order to check the compatibility of the various measurements with 
the results
of the fit, we make a comparison of the fit prediction obtained without using the observable of interest
as an input and the experimental measurement. 
Such predictions for a subset of observables are collected in 
Table~\ref{tab:predictions}.

\begin{table}
  \caption{Fit predictions obtained without including the corresponding
           experimental constraints into the fit itself. We quote the $68\%$ [$95\%$] probability
           ranges.}
  \label{tab:predictions}
  \begin{tabular}{|lll|}
    \hline
      $\alpha (^\circ)$ & $92.5\pm 4.2$ & [$84.3,100.5$]\\
      $\sin2\beta $ & $0.735\pm 0.034$  & [$0.672,0.800$]\\
      $\gamma (^\circ)$ & $64.4\pm 3.4$ & [$57.6,71.3$]\\
    \hline
      $|V_{ub}|\cdot 10^{3}$ & $3.48\pm 0.16$ & [$3.17,3.80$]\\
    \hline
      $\Delta m_{s} (\mathrm{ps}^{-1})$ & $17.0\pm 1.6$ & [$14.0,20.3$]\\
      $\sin2\beta_s$ & $0.0365\pm 0.0015$ & [$0.0337,0.0394$]\\
    \hline
  \end{tabular}
\end{table}

\begin{figure}
\centering
\includegraphics[width=.48\textwidth]{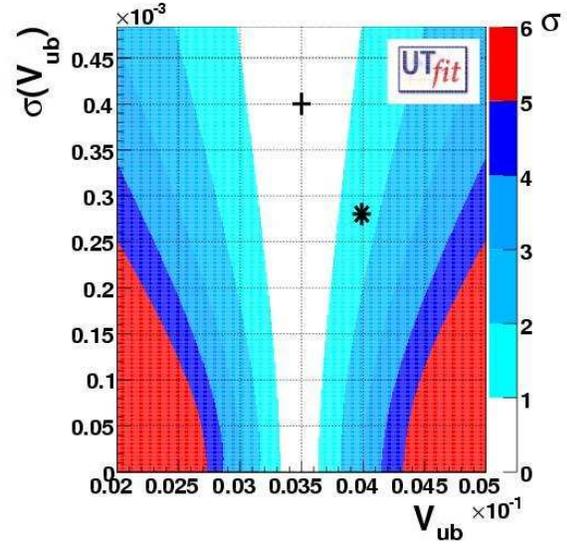}
\includegraphics[width=.48\textwidth]{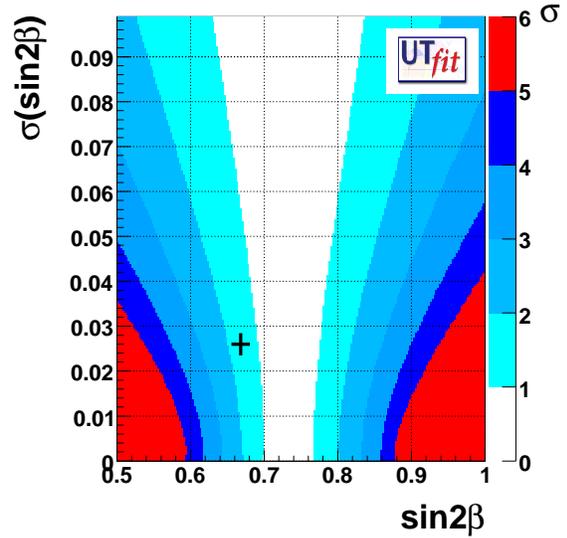}
\caption{Compatibility plots for $|V_{ub}|$ (left) and $\sin2\beta$ (right). 
The average value of the measurement is plotted on the horizontal axis, while its error 
is on the vertical one. The coloured bands delimit regions of values and errors 
which are less than a given number of $\sigma$ from the fit result. For $|V_{ub}|$, the exclusive 
(denoted by ``$+$'') and inclusive (denoted by ``$\ast$'') measurements are 
shown separately.}
\label{fig:compatibility}
\end{figure}

The two most significant discrepancies between measurements and fit predictions concern
$\sin2\beta$ and the inclusive determination of $|V_{ub}|$. As can be seen in
fig.~\ref{fig:compatibility}, they are at the level of $\sim1.5\sigma$, 
showing the excellent overall compatibility of the measurements with the 
SM fit (with the remarkable exception of the $B_s$ mixing phase, as we will see in the following).

The measured value of $\sin2\beta$ is $1.5\sigma$ smaller than the fitted one. Comparing with the results
of refs.~\cite{soni,guadagnoliburas}, we find that the SM fit using constraints from $|V_{ub}|$, $\varepsilon_K$ and $\Delta m_s$/$\Delta m_d$ 
only is again $1.5\sigma$ larger than the measurement, using the input values of Table~\ref{tab:inputs}. 

\section{The UT fit beyond the SM}
Once it is established that the CKM mechanism is the main source of CP violation so far, an 
accurate model-independent determination of $\bar\rho$ and $\bar\eta$ is extremely important for identifying NP in
 the flavour sector.

The generalized UT fit, using only $\Delta F=2$ processes and parametrizing generic NP
contributions, allows for the model-independent determination of $\bar\rho$ and $\bar\eta$ under the assumptions 
of negligible tree-level NP contributions. Details of the method 
can be found in ref.~\cite{utfitnp}.

A peculiar prediction of the SM
is that CP violation in $B_s$ mixing should be very small.
For this reason, the experimental observation of a sizable CP violation
in $B_s$ mixing would be an unambiguous signal of NP.

In fact, the present data give a hint of a $B_s$ mixing phase much
larger than expected in the SM, with a significance at about
$3\sigma$~\cite{utfitbetas}.
This result is obtained by combining all available experimental information
with the method used by our collaboration for UT analyses.

We perform a model-independent analysis of NP contributions to $B_s$
mixing using the following parameterization~\cite{utfiteft}:
\begin{eqnarray} C_{B_s}
  \, e^{2 i \phi_{B_s}} &=&\frac{A^\mathrm{SM}_s e^{-2 i \beta_s} +
    A^\mathrm{NP}_s e^{2 i (\phi^\mathrm{NP}_s - \beta_s)}}{A^\mathrm{SM}_s
    e^{-2 i \beta_s}} = \nonumber\\
 &=&\frac{\langle
    B_s|H_\mathrm{eff}^\mathrm{full}|\bar{B}_s\rangle} {\langle
    B_s|H_\mathrm{eff}^\mathrm{SM}|\bar{B}_s\rangle}\,, \quad 
  \label{eq:paranp}\nonumber
\end{eqnarray}
where $H_\mathrm{eff}^\mathrm{full}$ is the effective Hamiltonian
generated by both SM and NP, while $ H_\mathrm{eff}^\mathrm{SM}$ only
contains SM contributions. The angle $\beta_s$ is defined as $\beta_s=\arg(-(V_{ts}V_{tb}^*)/(V_{cs}V_{cb}^*))$ and it equals $0.018 \pm 0.001$ in the SM.

\begin{figure*}[t]
\begin{center}
\includegraphics[width=0.3\textwidth]{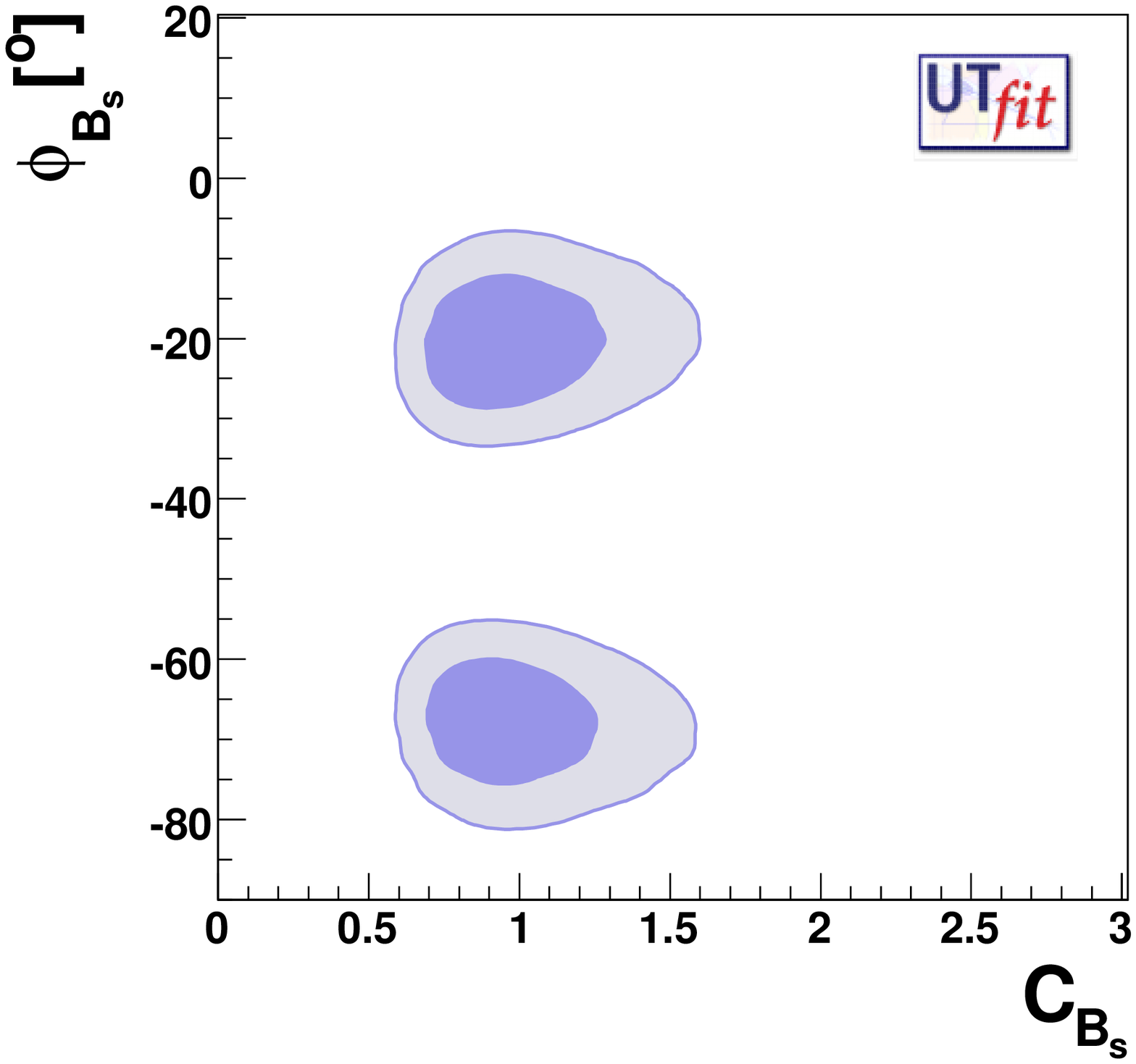}
\includegraphics[width=0.3\textwidth]{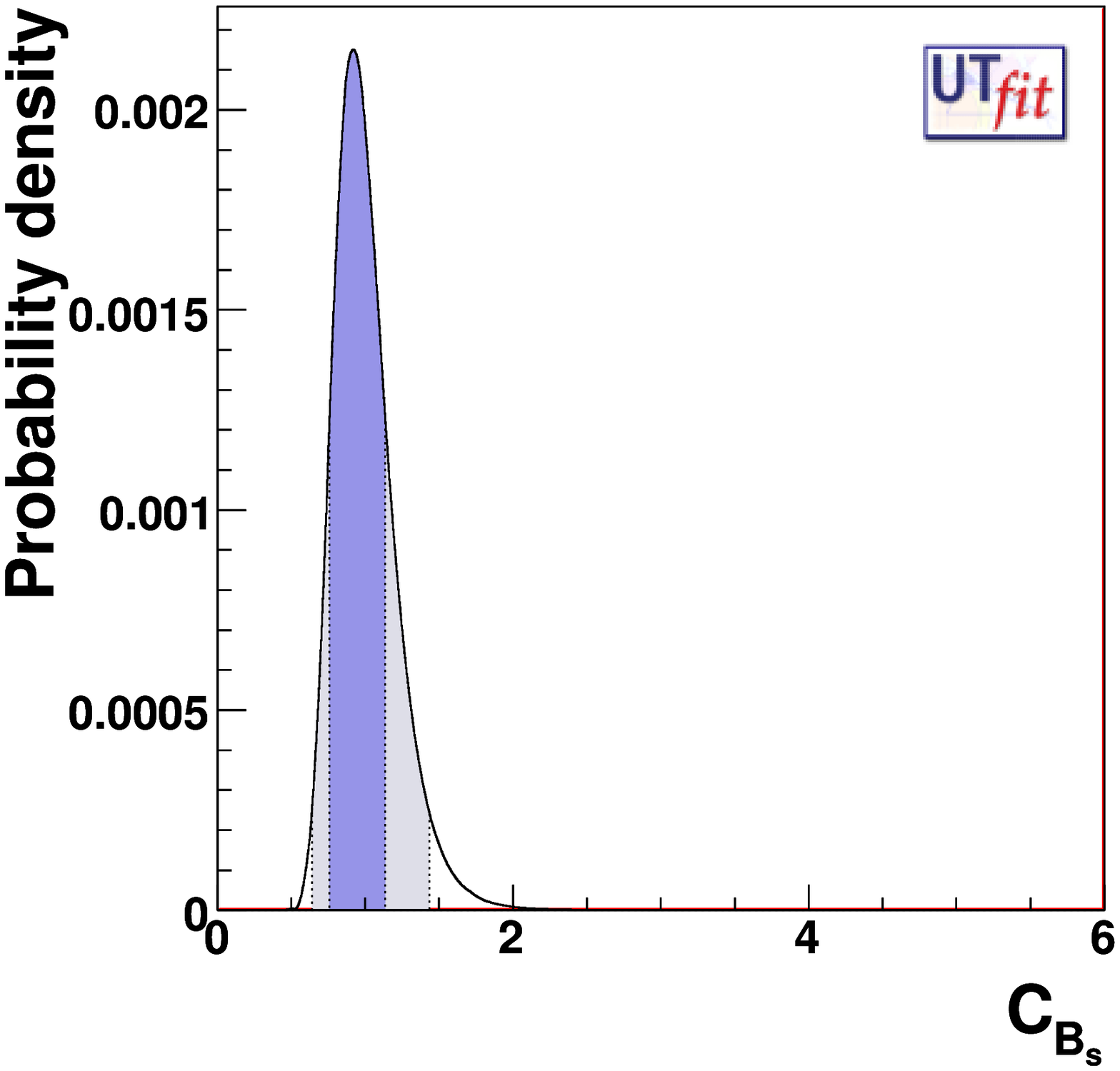}
\includegraphics[width=0.3\textwidth]{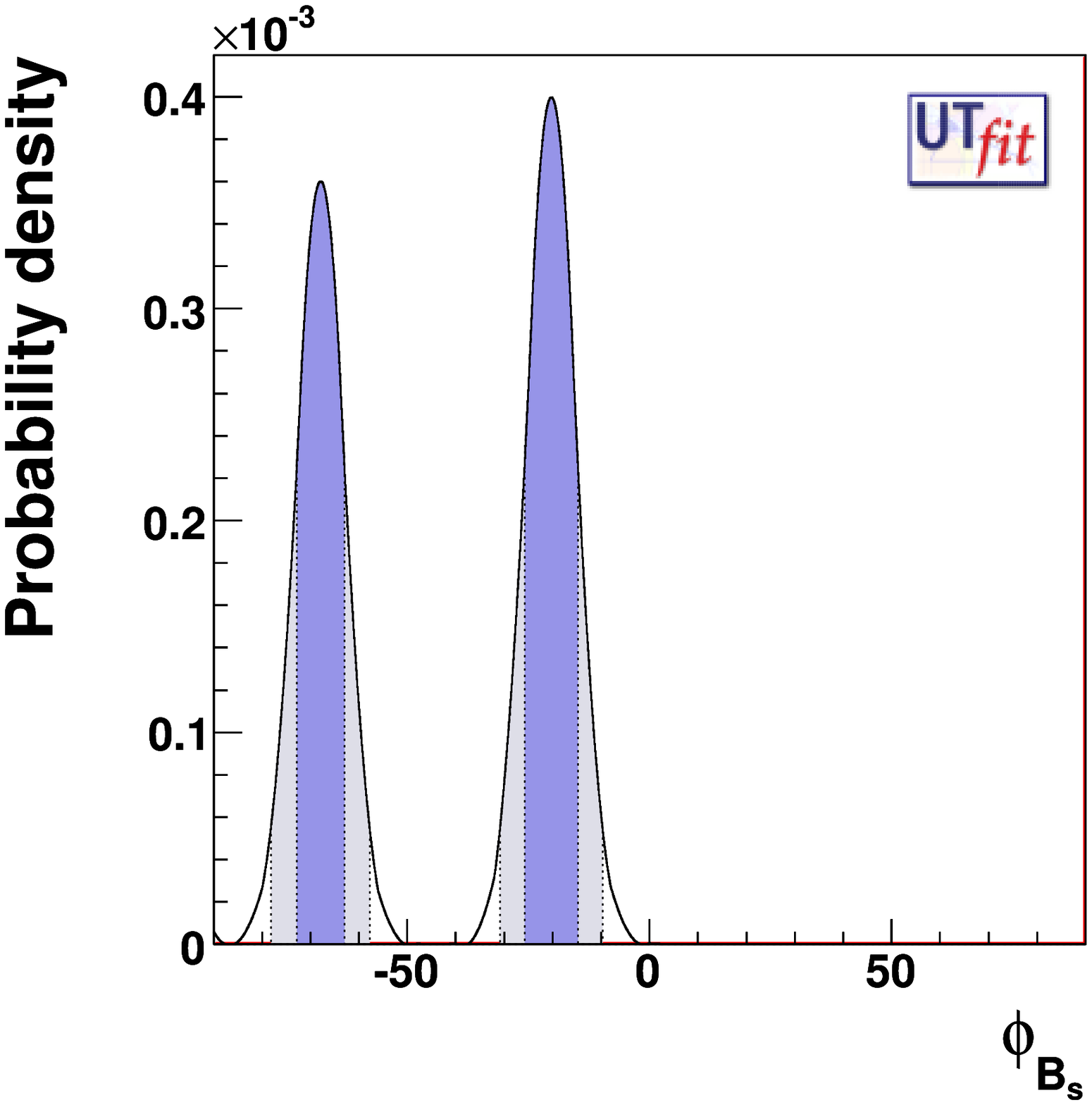}
\includegraphics[width=0.3\textwidth]{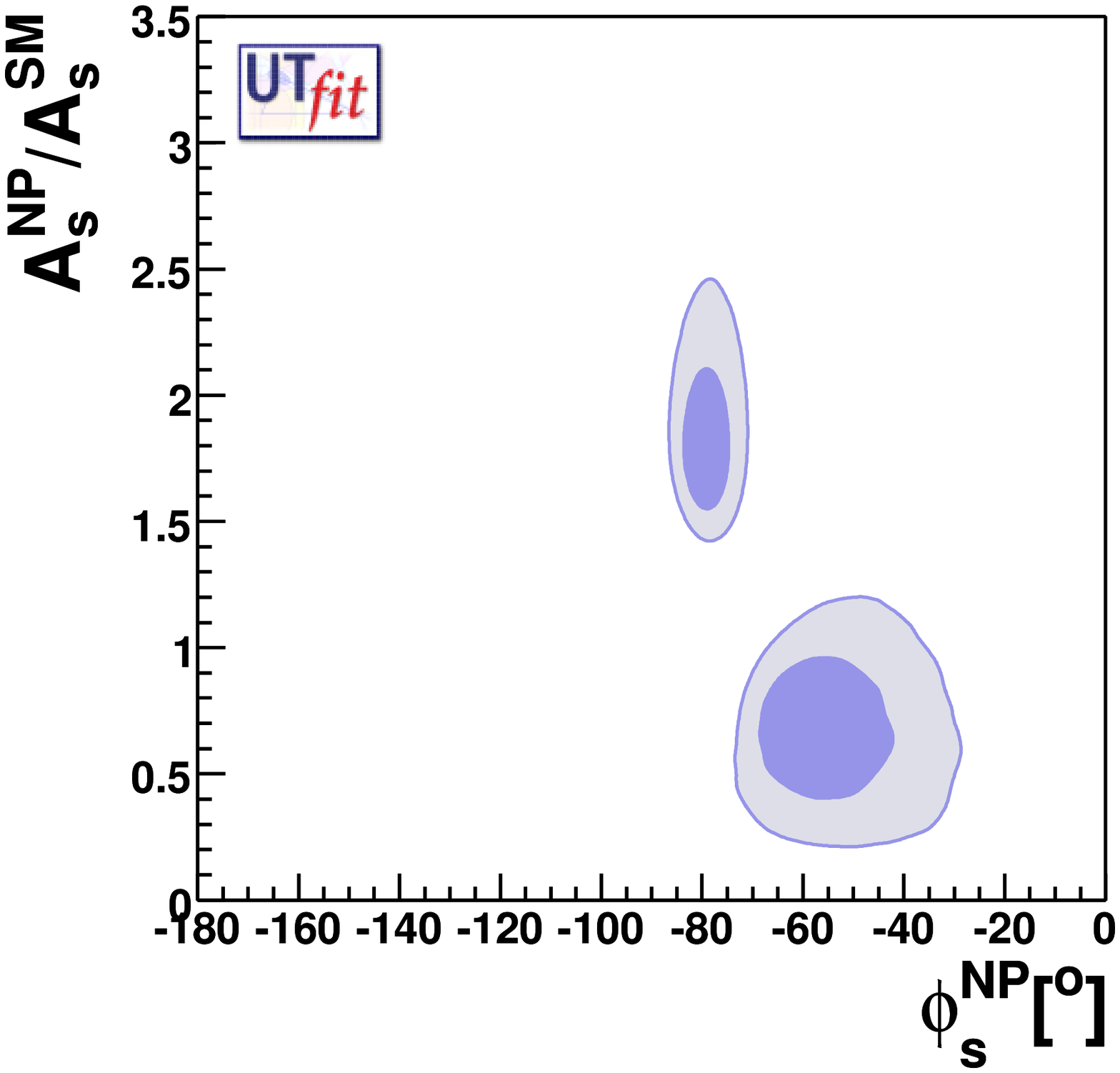}
\includegraphics[width=0.3\textwidth]{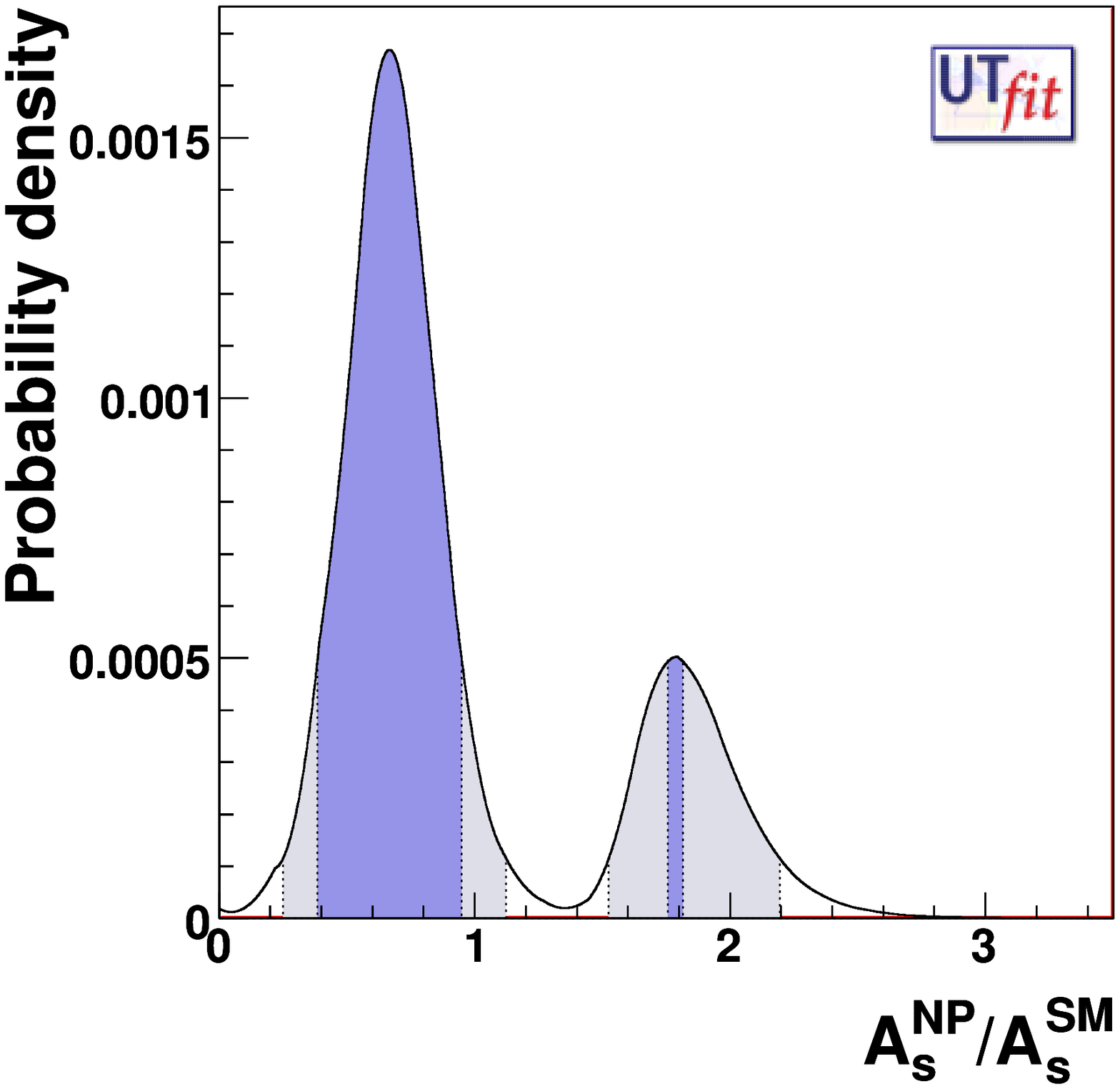}
\includegraphics[width=0.3\textwidth]{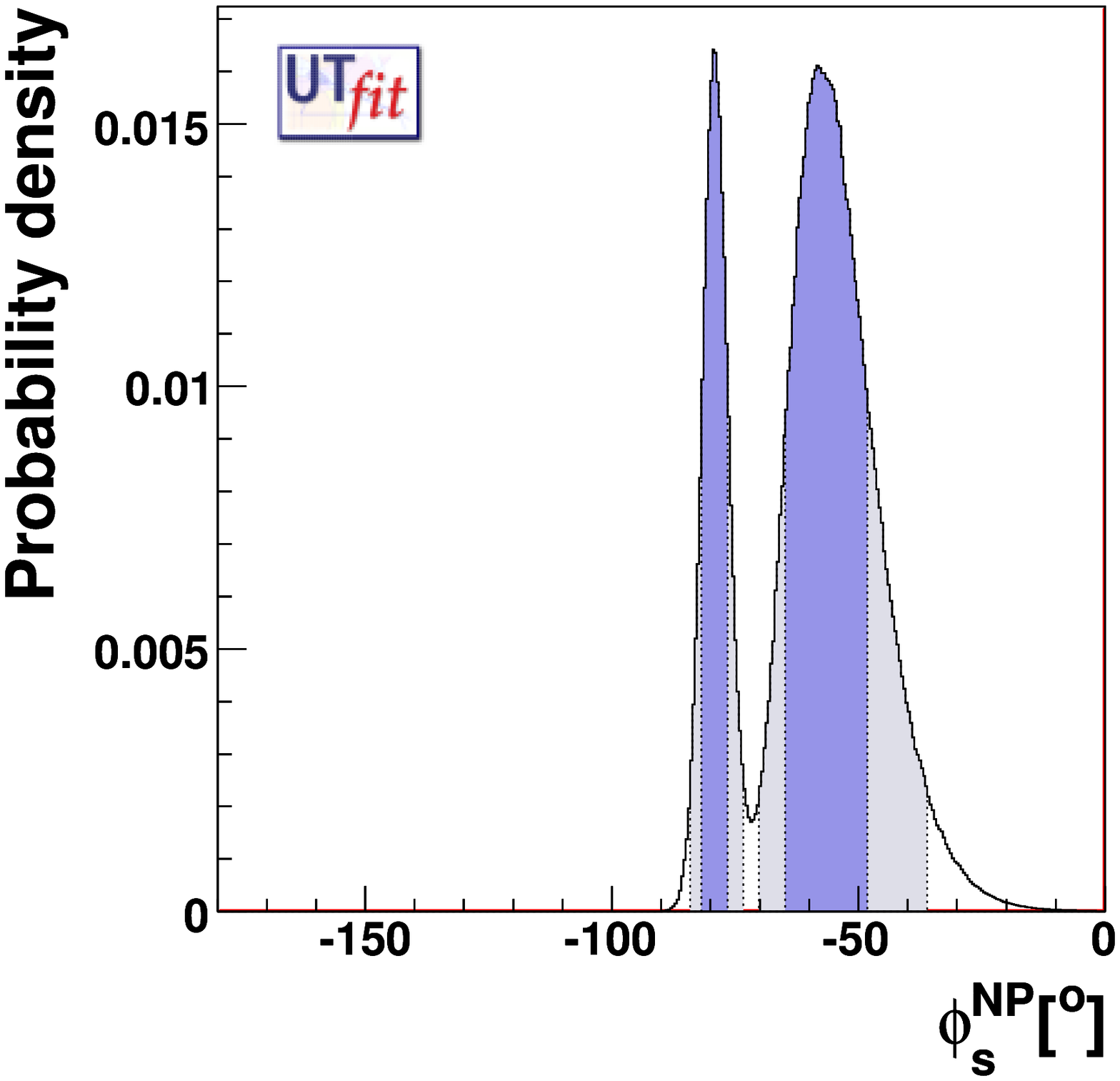}
\caption{%
  From left to right and from top to bottom: $68\%$ (dark) and  $95\%$
(light) probability regions in the $\phi_{B_s}$--$C_{B_s}$ plane;
p.d.f. for  $C_{B_s}$; p.d.f. for $\phi_{B_s}$; $68\%$ and $95\%$ probability regions 
in the $A^\mathrm{NP}_s/A^\mathrm{SM}_s$--$\phi^\mathrm{NP}_s$ plane;
p.d.f. for $A^\mathrm{NP}_s/A^\mathrm{SM}_s$; p.d.f. for $\phi^\mathrm{NP}_s$.}
\label{fig:NP}
\end{center}

\end{figure*}

We make use of the following experimental inputs: the CDF measurement of
$\Delta m_s$~\cite{dmsCDF}, the semi-leptonic asymmetry in $B_s$ decays
$A_\mathrm{SL}^{s}$~\cite{ASLD0}, the di-muon charge asymmetry
$A_\mathrm{SL}^{\mu\mu}$ from D\O~\cite{ACHD0} and
CDF~\cite{ASLCDF}, the measurement of the $B_s$ lifetime from
flavour-specific final states~\cite{tauBsflavspec}, the
two-dimensional likelihood ratio for $\Delta \Gamma_s$ and
$\phi_s=2(\beta_s-\phi_{B_s})$ from the time-dependent tagged angular
analysis of $B_s\to J/\psi \phi$ decays by CDF~\cite{CDFTAGGED} and
the correlated constraints on $\Gamma_s$, $\Delta \Gamma_s$ and
$\phi_s$ from the same analysis performed by D{\O}~\cite{D0TAGGED}.
For the latter, since the complete likelihood
is not available yet, we start from the results of the $7$-variable
fit in the free-$\phi_s$ case from Table I of ref.~\cite{D0TAGGED}. We
implement the $7 \times 7$ correlation matrix and integrate over the
strong phases and decay amplitudes to obtain the reduced $3 \times 3$
correlation matrix used in our analysis. In the D{\O} analysis, the
twofold ambiguity inherent in the measurement ($\phi_s \to \pi-\phi_s$, $\Delta \Gamma_s \to -\Delta \Gamma_s$, $\cos \delta_{1,2} \to - \cos \delta_{1,2}$) 
for arbitrary strong phases was
removed using a value for $ \cos \delta_{1,2}$ derived from the BaBar
analysis of $B_d \to J/\Psi K^*$ using SU(3). However, this neglects
the singlet component of $\phi$ and, although the sign of
$\cos \delta_{1,2}$ obtained using SU(3) is consistent with the
factorization estimate, to be conservative we reintroduce the ambiguity
in the D{\O} measurement, taking the errors quoted by D{\O} as Gaussian
and duplicate the likelihood at the point obtained by applying the
discrete ambiguity. Hopefully D{\O} will present results without
assumptions on the strong phases in the future, allowing for a more
straightforward combination. Finally, for the CKM parameters we
perform the UT analysis in the presence of arbitrary NP as described
in ref.~\cite{utfiteft}, obtaining $\overline {\rho} = 0.141 \pm 0.036$
and $\overline {\eta} = 0.373 \pm 0.028$.

\begin{table}[t]
\caption {Fit results for NP parameters, semi-leptonic asymmetries and
width differences. Whenever present, we list the two solutions due to the
ambiguity of the measurements. The first line corresponds to the one
closer to the SM.}

\label{tab:results2}
\begin{tabular}{|ccc|}
\hline
Observable  & $68\%$ Prob. & $95\%$ Prob.  \\
\hline
$\phi_{B_s} [^\circ]$             & -20.3 $\pm$ 5.3 & [-30.5,-9.9]\\
                                  & -68.0 $\pm$ 4.8 & [-77.8,-58.2]\\
\hline
$C_{B_s}$                         & 1.00 $\pm$ 0.20 & [0.68,1.51]\\
\hline
$\phi^\mathrm{NP}_s  [^\circ]$    & -56.3 $\pm$ 8.3 & [-69.8,-36.0]\\
                                  & -79.1 $\pm$ 2.6 & [-84.0,-72.8]\\
\hline
$A^\mathrm{NP}_s/A^\mathrm{SM}_s$  & 0.66 $\pm$ 0.28 & [0.24,1.11]\\
                                   &  1.78 $\pm$ 0.03 & [1.53,2.19]\\
\hline
\end{tabular}
\end{table}

The results of our analysis are summarized in Table \ref{tab:results2}.
We see that the phase $\phi_{B_s}$ deviates from zero at more than $3.0\sigma$.
In Fig.~\ref{fig:NP} we present the two-dimensional
$68\%$ and $95\%$ probability regions for the NP parameters $C_{B_s}$
and $\phi_{B_s}$, the corresponding regions for the parameters
$A^\mathrm{NP}_s/A^\mathrm{SM}_s$ and $\phi^\mathrm{NP}_s$, and the
one-dimensional distributions for NP parameters.

The solution around $\phi_{B_s} \sim -20^\circ$ corresponds to
$\phi^\mathrm{NP}_s \sim -56^\circ$ and $A^\mathrm{NP}_s/A^\mathrm{SM}_s \sim 79\%$.
The second solution is much more distant from the SM and it requires a
dominant NP contribution ($A^\mathrm{NP}_s/A^\mathrm{SM}_s \sim 180\%$)
and in this case the NP phase is very well determined.

Finally, we have tested the significance of the NP signal against different
modeling of the probability density function (p.d.f.). We have explored
two more methods with respect to the standard Gaussian one used by the
D{\O} Collaboration in presenting the result: this is mainly to address the
non-Gaussian tails that the experimental likelihood is showing.
Firstly, we have used the $90\%$ C.L. range for
$\phi_s=[-0.06,1.20]^\circ$ given by D{\O} to estimate the standard
deviation, obtaining $\phi_s=(0.57 \pm 0.38)^\circ$ as input for the
Gaussian analysis. This is conservative
since the likelihood has a visibly larger half-width on the side
opposite to the SM expectation (see Fig.~2 of Ref.~\cite{D0TAGGED}).
Second, we have implemented the likelihood profiles for $\phi_s$ and
$\Delta \Gamma_s$ given by D{\O}, discarding the correlations but
restoring the strong phase ambiguity.  The likelihood profiles include
the second minimum corresponding to $\phi_s \to \phi_s + \pi$, $\Delta \Gamma \to - \Delta \Gamma$, 
which is disfavoured by the oscillating
terms present in the tagged analysis and is discarded in the Gaussian
analysis. Also this approach is conservative since each
one-dimensional profile likelihood is minimized with respect to the
other variables relevant for our analysis. It is remarkable that both
methods give a deviation of $\phi_{B_s}$ from zero of $3\, \sigma$.
We conclude that the combined analysis gives a stable
departure from the SM, although the precise number of standard deviations
depends on the procedure followed to combine presently available data.

\end{document}